# Adaptive Weighted Genetic Algorithm-Optimized SVR for Robust Long-Term Forecasting of Global Stock Indices for investment decisions

*Mohit Beniwal*

*Delhi School of Management, Delhi Technological University*

ARTICLE INFO

*Article history:*
Received
Received in revised form
Accepted

ABSTRACT

Long-term price forecasting remains a formidable challenge due to the inherent uncertainty over the long term, despite some success in short-term predictions. Nonetheless, accurate long-term forecasts are essential for high-net-worth individuals, institutional investors, and traders. The proposed improved genetic algorithm-optimized support vector regression (IGA-SVR) model is specifically designed for long-term price prediction of global indices. The performance of the IGA-SVR model is rigorously evaluated and compared against the state-of-the-art baseline models, the Long Short-Term Memory (LSTM), and the forward-validating genetic algorithm optimized support vector regression (OGA-SVR). Extensive testing was conducted on the five global indices, namely Nifty, Dow Jones Industrial Average (DJI), DAX Performance Index (DAX), Nikkei 225 (N225), and Shanghai Stock Exchange Composite Index (SSE) from 2021 to 2024 of daily price prediction up to a year. Overall, the proposed IGA-SVR model achieved a reduction in MAPE by 19.87% compared to LSTM and 50.03% compared to OGA-SVR, demonstrating its superior performance in long-term daily price forecasting of global indices. Further, the execution time for LSTM was approximately 20 times higher than that of IGA-SVR, highlighting the high accuracy and computational efficiency of the proposed model. The genetic algorithm selects the optimal hyperparameters of SVR by minimizing the arithmetic mean of the Mean Absolute Percentage Error (MAPE) calculated over the full training dataset and the most recent five years of training data. This purposefully designed training methodology adjusts for recent trends while retaining long-term trend information, thereby offering enhanced generalization compared to the LSTM and rolling-forward validation approach employed by OGA-SVR, which forgets long-term trends and suffers from recency bias.

## 1. Introduction

A stock market plays an important role in the financial system of a country [1]. A country needs a robust financial system, and the stock market can be regarded as an indicator. The popularity of the stock market increased tremendously in India [2] in the last five years, from 2020 to 2025. Investors are generally interested in markets that are experiencing growth. However, investing carries many risks, and the significant volatility observed in the stock market during the COVID-19 pandemic has further compounded the uncertainty. Investors' motivation is also influenced by academic theories such as the Efficient Market Hypothesis [3] and Random Market Hypothesis [4]. The Efficient Market Hypothesis claims that investors cannot obtain excess returns from the market over the long term. In short, financial markets are unpredictable, a conclusion that the Random Walk Hypothesis also supports. However, the predictability of the stock market is inconclusive. The Inefficient market hypothesis states that markets are not always efficient [5]. Numerous studies challenge the claims of the Efficient Market Hypothesis and the Random Walk Hypothesis [6–10]. Benjamin Graham emphasized the dual nature of the stock market, which functions as a voting machine in the short term, whereas, in the long term, it operates as a weighing machine [11].

There are several approaches to forecast the stock market, including fundamental analysis [12], which focuses on financial statements and macroeconomic indicators; technical analysis [13], which analyzes historical price patterns and trading volumes; quantitative techniques, such as statistical and econometric methods [14–16]; machine learning and artificial intelligence algorithms that utilize large datasets for pattern recognition [17]; and sentiment analysis, which interprets market sentiment from news, social media, and investor behavior. The advent of Artificial Intelligence (AI) and Machine Learning (ML) has significantly impacted financial markets [18], especially after the rise of AI tools like ChatGPT, the focus on artificial intelligence has taken center stage [19].

ML can be categorized into supervised and unsupervised learning algorithms [20]. Supervised learning algorithms rely on labeled data, where each input is paired with a corresponding output label [21]. The algorithm maps the input features and their corresponding output labels, which enables it to predict unseen data. This methodology can be applied to both classification and regression. Unsupervised learning algorithms, on the other hand, can identify hidden patterns using unlabeled data and are commonly used in tasks such as clustering, dimensionality reduction, and anomaly detection. Some popular supervised learning algorithms are Support Vector Machine/Regression (SVM/SVR) and Artificial Neural Networks (ANNs) [22]. Deep learning (DL), a specialized subset of ML, can capture complex patterns from large datasets [23,24]. Recurrent Neural Networks (RNNs), Convolutional Neural Networks (CNNs), Gated Recurrent Units (GRUs), Long Short-Term Memory (LSTM) networks, and Bidirectional Long Short-Term Memory (Bi-LSTM) networks are some widely used deep learning algorithms [25–28]. LSTM and SVM/SVR are the most commonly used ML algorithms for forecasting financial time series [29–31]. This study employs Support Vector Regression (SVR) and Long Short-Term Memory (LSTM) algorithms to forecast the stock indices. Figure 1 shows the classification of AI [32].

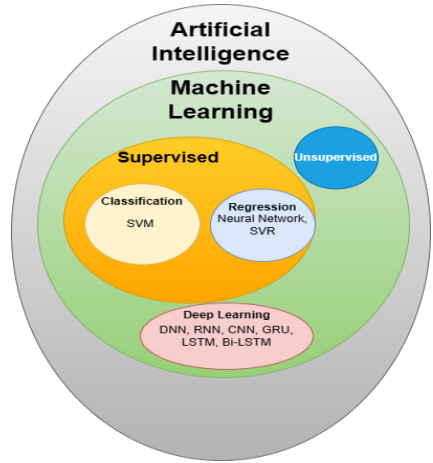

**Figure 1.** Classification of AI

Accurate forecasting helps investors make informed decisions [33] and mitigate risk, but the noisy nature of financial time series makes it a daunting task [34,35]. Market anomalies disrupt patterns and further complicate forecasting financial time series [32]. Developing a forecasting model capable of handling diverse financial time series from different economies is a significant challenge. While most studies forecast one-day-ahead stock prices, some predict intraday prices [36], which is useful for frequent trading. In their literature review, Rouf et al. [37] reviewed 24 studies, and most of them forecasted short-term prices, such as daily or intraday, but none predicted long-term prices. Investors are also interested in long-term price forecasts, such as daily price patterns up to a year. Although yearly frequency is commonly used, it fails to capture price patterns.

Financial time series using ML poses a series of challenges to accurate predicting [38,39]. ML algorithms require careful tuning of hyperparameters, and manual parameter selection may introduce biases into the research [40]. A common algorithm for automating hyperparameter optimization is the Genetic Algorithm (GA) [41]. Genetic algorithms are inspired by natural selection and genetics [42] to find the optimum hyperparameters [43]. SVR has three important hyperparameters: C, epsilon (ε), and gamma for non-linear kernels. C balances model complexity and training accuracy, ε sets a margin where small errors are ignored, and gamma controls how much influence each data point has in nonlinear kernels. In this study, genetic algorithms are utilized to tune these hyperparameters of the SVR model. Forward validation is a widely employed approach for mitigating bias in time series forecasting. This method may prevent look-ahead bias and the possible leakage of future information into the training dataset. Schnaubelt [44] empirically studied common validation schemes and concluded that forward validation provides more accurate estimates. The rolling forward validation model is trained on earlier windows and subsequently evaluated on later ones, a design suitable for financial time series data. The rolling window forward validation procedure captures inherent temporal patterns; however, this approach may also induce recency bias by focusing on rolling windows, potentially forgetting long-term patterns in financial time series. In contrast, the proposed model leverages the entire training dataset of financial time series for training, thereby preserving long-term trend information. Furthermore, the scoring metrics during training also assign a higher weightage to recent data to better adapt to current trends. For comparative purposes, state-of-the-art baseline models, such as the rolling forward validating genetic algorithm (OGA-SVR) proposed by Beniwal et al. [45] and the Long Short-Term Memory (LSTM) are employed.

This study comprehensively evaluates the same five global indices used by Beniwal et al. [45] across a multi-year testing window from 2021 to 2024. Hence, the study examines India's Nifty, the United States' Dow Jones Industrial Average (DJI), Germany's DAX Performance Index (DAX), Japan's Nikkei 225 (N225), and China's Shanghai Stock Exchange Composite Index (SSE). These global indices are also the top five economies of the world in terms of Gross Domestic Product (GDP) [46]. The study proposes the Improved Genetic Algorithm-optimized Support Vector Regression (IGA-SVR) model with robust long-term forecasting performance utilizing a data-driven selection of hyperparameters during training. A series of experiments were conducted to optimize hyperparameters and evaluate training approaches, resulting in an improved model. The IGA-SVR model was trained on the entire dataset, with additional weightage assigned to recent training data. For future predictions, the final selection of optimized parameters using genetic algorithms from training was based on the arithmetic mean of the MAPE computed over both the full training dataset and five-year recent training data.

The objectives of this study are:

- Conducts extensive evaluation of IGA-SVR, LSTM, and OGA-SVR models on five global stock indices over a yearly testing period from 2021 to 2024.
- Examines the training approach used by IGA-SVR, which leverages the entire training dataset while giving more weightage to recent training data, compared to rolling forward validation used by OGA-SVR.
- Demonstrate the superior performance of the proposed Improved Genetic Algorithm-Optimized Support Vector Regression (IGA-SVR) model over state-of-the-art models.

The study provides an improved model for investors to make insightful decisions before investing in global stock indices. The remainder of the paper is organized as follows: Section 2 presents the literature review. Section 3 provides the basic concepts of the standard algorithms. Section 4 discusses methodology which explains the prediction process and the IGA-SVR model. Section 5 reports and discuss the results. Finally, Section 6 concludes the study, suggests the scope for future research, and highlights the limitations of the study.

## 2. Literature Review

AutoRegressive Integrated Moving Average (ARIMA) is one of the most frequently used statistical methods for time series analysis [47]. However, ARIMA is generally not preferred in predicting financial time series, especially for the long term [34,48]. Recently, AI has been in the limelight, especially after the advent of ChatGPT, Gemini, and Deepseek. There is interest in the researchers to improve the prediction of financial time series. Artificial Neural Network (ANN) is inspired by biological neurons and is frequently used for complex pattern recognition. Teixeira Zavadzki de Pauli [49] applied ANNs to predict Brazilian stocks. Their result indicated that the performance of the ANNs was satisfactory. However, ANNs do not retain the sequential information present in time series data. Recurrent Neural Networks (RNNs) have special architecture to handle sequential data. Yang et al. [50] proposed a hybrid model that integrated an enhanced Particle Swarm Optimization (PSO) algorithm with RNN to achieve superior stock price prediction accuracy. One problem with RNNs is that they are susceptible to the vanishing gradient phenomenon, which degrades their ability to learn long-range dependencies [51,52]. LSTMs are specifically designed to mitigate the vanishing gradient problem [53,54]. Mehtab et al. [55] concluded that the LSTM-based model was the most accurate among all the models tested in their study. Chen et al. [56] tested their LSTM models on China's commercial bank stocks and concluded that their model had superior generalization ability. Kumar and Haider [57] experimented with a hybrid RNN-LSTM model utilizing metaheuristic optimization techniques. Their results showed that the metaheuristic approach increased the forecasting accuracy. Md et al. [58] asserted that their novel model based on Multi-layer sequencial LSTM gave superior forecasting performance than other evaluated ML and DL algorithms. Lu et al. [59] proposed a hybrid model based on bi-directional LSTM (BiLSTM), which demonstrated superior performance compared to the other models in that study.

Another frequently used ML algorithm for stock prediction is SVM/SVR, which has been effective for forecasting [60]. SVM/SVR has also shown excellent performance in predicting stock prices [61]. SVM/SVR is a kernel-based algorithm that utilizes hyperplanes in a higher-dimensional space for class separation or prediction. SVM is applied to classification tasks, whereas SVR is employed for regression tasks [62]. Zou et al. [63] proposed a Twin Support Vector Machine model and used technical indicators as input features to forecast the next day's stock price direction. Their results indicated that the proposed model outperformed other benchmark models. Xiao et al. [64] introduced a hybrid ARI-MA-LS-SVM model that combined cumulative auto-regressive (AR) moving average (MA) with a least squares (LS) support vector machine (SVR), showing increased accuracy and market applicability. While LSTM can retain sequential information in time series data [65], SVM/SVR does not possess such an inbuilt mechanism. SVM/SVR models require effective optimization techniques to fine-tune hyperparameters [66]. Commonly used approaches for this purpose include Grid Search (GS) and the Genetic Algorithm (GA) [67]. Optimizing hyperparameters of SVM/SVR minimizes training error and reduces the risk of overfitting [68], thereby enhancing their generalization ability. Mahmoodi et al. [69] used PSO with SVM to improve the classification. Dash et al. [70] introduced a fine-tuned Support Vector Regression using a grid search to optimize the performance of SVR. They concluded that their approach was faster and more accurate in predicting stock performance.

Genetic algorithms are a population-based metaheuristic inspired by evolutionary processes and have been applied to a variety of applications [71]. Li and Sun [72] presented a predictive model combining kernel parameters and parameter optimization based on the SVM model. GA can also be used to find optimal configurations for deep learning models. Gao et al. [73] proposed a deep learning model optimized using GA to forecast the overnight return direction of the stock market indices. Solares et al. [74] evaluated a system combining GA and ANN using the S&P 500 index and compared its performance against multiple benchmarks. Their results showed that the proposed system outperformed the benchmarks. Thakkar and Chaudhari [75] introduced a GA-based information fusion approach to optimize the parameter section and feature extraction of LSTM. Their proposed method was superior to existing GA-based optimization techniques.

### 2.1. Research Gaps

The literature review suggests that only a few studies focused on developing countries [76]. Research in this domain is generally divided between predictions for market indices and individual stocks [77]. This study evaluates the proposed model on global stock market indices from both developed and developing countries. Prediction model development is undergoing rapid change [78]. Previous studies in stock prediction have primarily focused on forecasting next-day prices or short-term prices. However, there has been a lack of emphasis on developing a model for long-term forecasting, an area of

Author at: Delhi School of Management, Delhi Technological University, Delhi 110042, India
E-mail addresses: mohitbeniwal@dtu.ac.in (M. Beniwal)

interest for investors in financial time series forecasting. This study addresses this gap by improving predictions of daily price patterns over a one-year horizon. Moreover, prior research has largely concentrated on optimizing the hyperparameters C and epsilon; optimizing the gamma parameter is ignored in Support Vector Regression (SVR) models, which have the Radial Basis Function (RBF) as the default kernel in the scikit-learn library. This study optimizes all three parameters, namely C, epsilon, and gamma. While most studies employ cross-validation and some utilize forward validation or rolling forward validation with SVR, these approaches fail to maintain sequential information. This study utilizes a purposefully designed methodology that is capable of utilizing long-term trends and adjusting to recent trends. Further, this study benchmarks the proposed model against the state-of-the-art model by Beniwal et al. [45]. Few studies evaluate their models on multiple global stock indices over multiple yearly prediction windows. This study comprehensively evaluates the performance of the proposed model on yearly predictions from 2021 to 2024.

## 3. Conceptual Overview

In this section, a brief overview of the basic concepts and workings of LSTM, SVR, and GA is provided.

### 3.1. Support Vector Regression (SVR)

The development of modern Support Vector Machine (SVM) is attributed to the seminal contributions of Vapnik et al. [79–84]. SVM falls under supervised ML and constructs one or more hyperplanes in a high-dimensional space, which can be utilized for classification [85]. SVM finds a hyperplane that maximizes the margin between support vectors of different classes [86]. Furthermore, SVM uses kernel functions and can transform non-linearly separable data into higher-dimensional spaces, where a linear decision boundary can effectively separate the classes. SVR, on the other hand, aims to approximate target values by not penalizing errors within the epsilon-insensitive tube while penalizing errors outside the tube. Figure 2 shows the SVM and SVR.

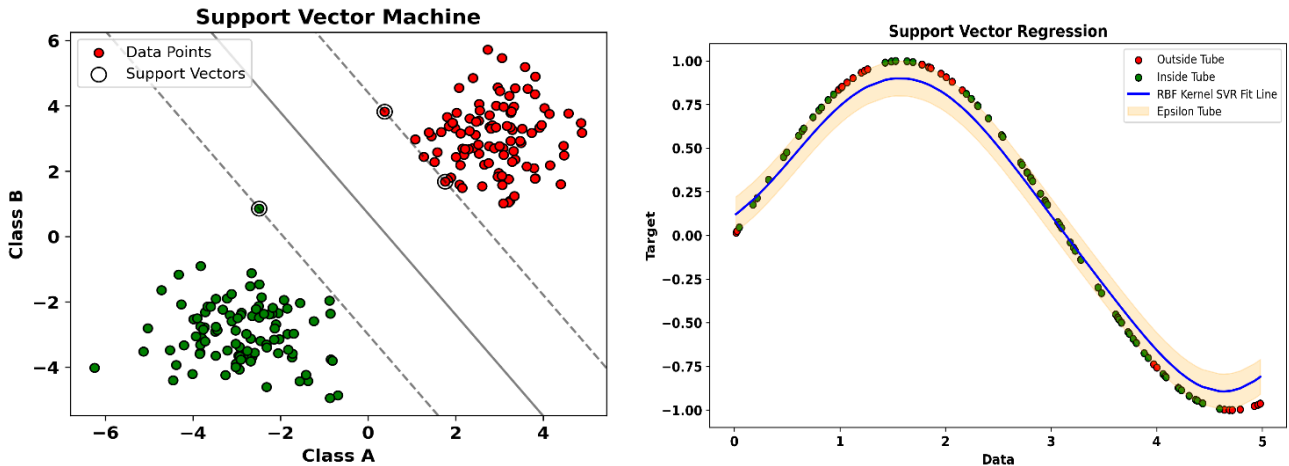

**Figure 2**. SVM and SVR

### 3.2. Long Short-Term Memory (LSTM)

Long Short-Term Memory (LSTM) was introduced by Hochreiter and Schmidhuber [87]. LSMT includes gating mechanisms and memory cells, helping address the vanishing gradient problem of RNNs. Due to this, LSTM has an excellent ability to capture long-term dependencies in a sequence and forgets unnecessary information. This makes LSTM well-suited for financial time series. The LSTM has three main gates, namely the input gate, output gate, and forget gate [88]. These gates selectively retain important information and forget unimportant information in time series data. Figure 3 shows the architecture of LSTM.

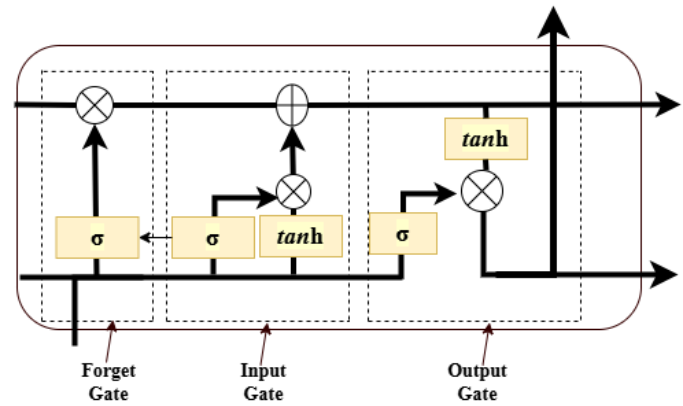

**Figure 3**. LSTM

### 3.3. Genetic Algorithm (GA)

Genetic Algorithms (GAs) are computational methods inspired by the theory of evolution proposed by Darwin [89,90]. They can explore a large solution space to find an optimum solution [91,92]. In SVR, GA can be used to identify the optimum value of hyperparameters. Generally, a GA involves six steps: initialization, fitness evaluation, stopping criteria, selection process, crossover, and mutation [93]. Initially, GAs generate random candidate solutions from a population of chromosomes. A fitness function is then used to evaluate and select the fittest solutions based on the objective of either minimizing or maximizing the function. Mostly approximate optimum solutions are selected, with the stopping criterion often defined by the number of generations. Figure 4 illustrates the Genetic Algorithm process.

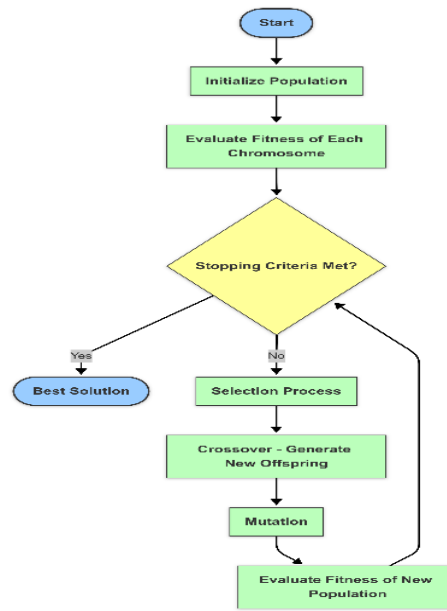

**Figure 4.** Genetic Algorithm

Author at: Delhi School of Management, Delhi Technological University, Delhi 110042, India
E-mail addresses: mohitbeniwal@dtu.ac.in (M. Beniwal)

## 4. Methodology

### 4.1. The Data

The data was acquired from Yahoo Finance. The indices data for the top five GDP countries [94] was used for analysis; these countries, in descending order of GDP, are the US, China, Germany, Japan, and India. The data was collected from April 2008, as this was the earliest common date when historical data for all indices was available. The data is taken till December 2024. Figure 5 shows the comparison of the cumulative returns of stock indices. Nifty has been the leading index in terms of return, followed by the Dow Jones Industrial Average (DJI), Nikkei 225 (N225), DAX Performance Index (DAX), and SSE Composite Index (SSE). It can be noted that the SSE Composite Index has shown almost no return over the past 15 years. Table 1 shows the positive and negative closing of indices.

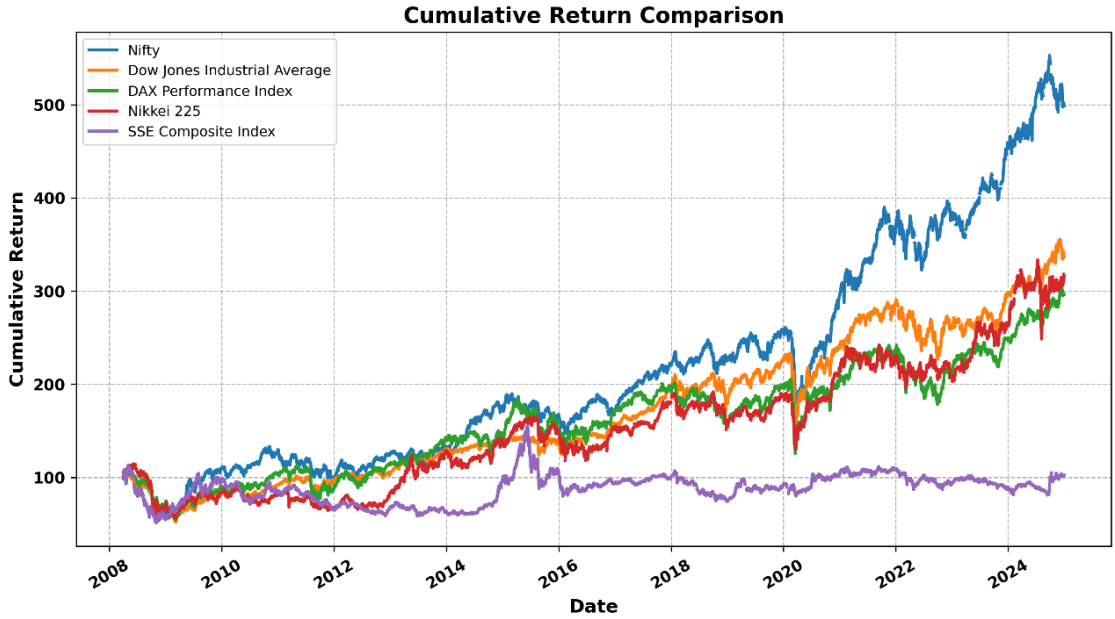

**Figure 5. Comparison of the Cumulative returns of stock indices**

**Table 1. Index Data Summary**

| Country Name | Index Name | Positive Closings | Negative Closings |
|---|---|---|---|
| USA | Dow Jones Industrial Average (DJI) | 54% | 46% |
| INDIA | NIFTY 50 (NIFTY) | 53% | 47% |
| GERMANY | Deutscher Aktienindex (DAX) | 53% | 47% |
| JAPAN | Nikkei 225 (N225) | 53% | 47% |
| CHINA | SSE Composite Index (SSE) | 52% | 48% |

*4.2. Prediction Process*

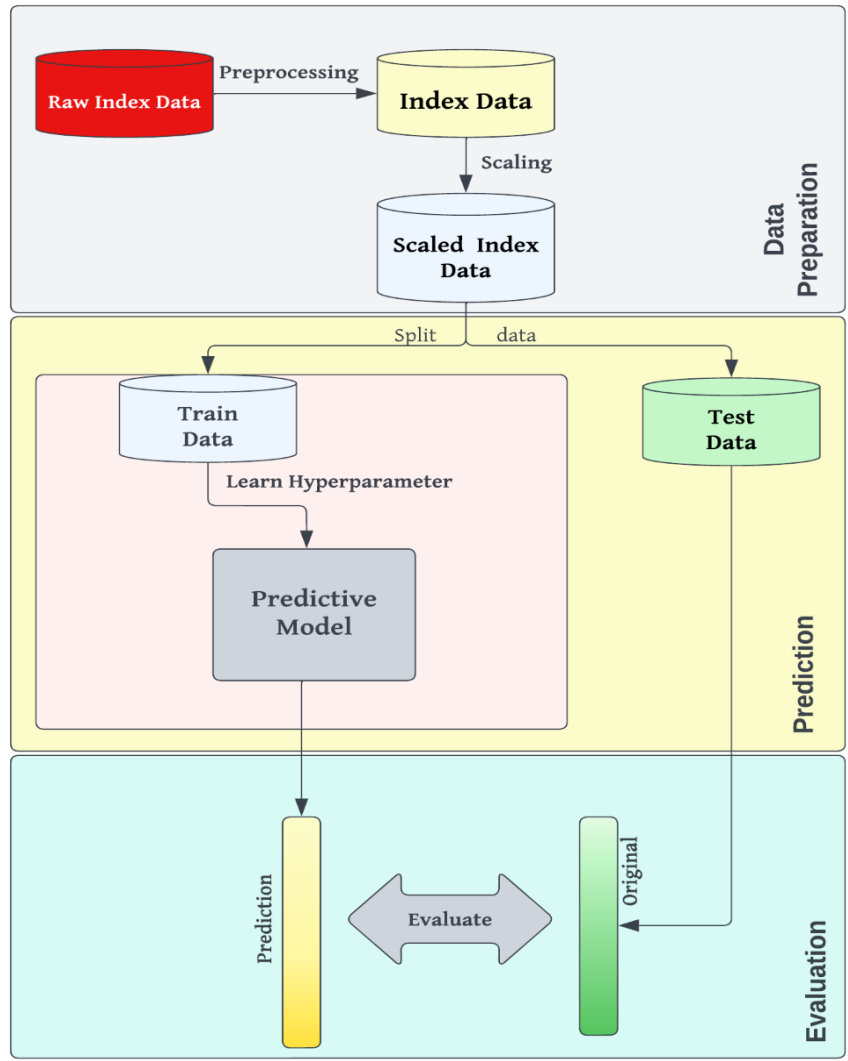

**Figure 6. Prediction Process**

Figure 6 shows the flowchart of the prediction process. First, the raw index data is fetched from Yahoo Finance and preprocessed to forward-fill any missing data or holidays specific to each country to ensure that all stock indices have the same number of data points. The preprocessed index data is scaled using a standard min-max scaler, which scales the date index and close prices from 0 to 1. This scaled data is then divided into training and testing datasets. The training data is used by the GA, which identifies an optimum hyperparameter, such as C, epsilon, and gamma. The optimum hyperparameters are subsequently used for forecasting using the SVR model. The predicted prices from the proposed model are compared with the testing price data to evaluate the performance.

Author at: Delhi School of Management, Delhi Technological University, Delhi 110042, India
E-mail addresses: mohitbeniwal@dtu.ac.in (M. Beniwal)

*4.3. Improved Genetic Algorithm Optimized Support Vector Regression (IGA-SVR)*

The Improved Genetic Algorithm-Optimized Support Vector Regression (IGA-SVR) model is trained on the entire training dataset, with higher weightage assigned to the most recent five years' close prices. Table 2 presents the pseudocode of the IGA-SVR prediction process. The dataset is divided into training and test datasets, with the last year of data reserved for testing, while the remaining data is used for training. Generally, GA searches for optimal solutions in a stochastic though guided manner. However, if the search space is unrestricted, it can lead to huge computation costs with the risk of finding suboptimal solutions by getting trapped in local optima. Therefore, logical restrictions are imposed on the hyperparameters of the SVR model and the number of generations. Based on the literature, the number of generations is restricted to 30 [95]. The SVR is implemented using the scikit-learn library. The default values of SVR hyperparameters in scikit-learn are: C = 1, epsilon = 0.1, and gamma = 'scale'. The 'scale' value is computed as $1/(n * variance of features)$, where n is the number of features in the dataset. The regularization parameter C controls the trade-off between model complexity and training error. A higher value of C may lead to overfitting, while a very low value may cause underfitting. In scikit-learn, C must be greater than zero. A moderate value of C tolerates some training error, which can improve generalization, important for noisy stock market time series. Hence, 0.01 is chosen as the minimum value, and the default value 1 is chosen as the maximum. Similarly, epsilon controls the margin of tolerance within which no penalty is given for errors. A small value of epsilon may lead to overfitting, while a higher epsilon increases tolerance, which may result in underfitting. Therefore, the default value of 0.1 in the scikit-learn library is taken as the minimum, and the maximum value is set to 1. For gamma, the minimum possible value is zero. Higher gamma makes the model too sensitive, so selecting an appropriate maximum value is challenging. A trial-and-error approach is adopted, where 'scale'/2, 'scale'/5, and 'scale'/10 are tested. Among these, 'scale'/10 is found to be the most suitable for long-term stock index prediction. Table 3 lists the restrictions imposed on the GA solution search space to identify appropriate values for the IGA-SVR hyperparameters. The configuration of baseline models OGA-SVR and LSTM was the same as defined in Beniwal et al. [45]. IGA-SVR has the following improvements compared to OGA-SVR:

- The IGA-SVR model is trained on the entire training dataset, allowing it to capture the long-term trend of the index. Additionally, the MAPE calculated from the most recent five years of training data is combined with the MAPE over the full training dataset using an arithmetic mean. This approach assigns a higher weight to recent data and selects a model that captures both the long-term trend and recent trend, thus improving model robustness.
- Maximum and minimum restrictions are placed on the SVR hyperparameters to ensure that the genetic algorithm searches within the defined boundaries of the search space. This also helps prevent the model from becoming overly sensitive to noise and improves generalization.
- The OGA-SVR model used rolling forward validation, which may fail to retain long-term trend information and can become biased toward recent trends due to the presence of recent data in the rolling window.

**Table 2. Pseudo code of IGA-SVR**

| Algorithm |
| --- |
| **Input**: Scaled Training Data |
| **Output**: Daily Close Price Prediction of 1 year |
|     **Preprocessing** |
|     Forward fill the downloaded data. |
|     Scale the data using a min-max scaler. |
|     **Loop** for multiple years |
|       Split the data, keeping 1 year of data as test data and the rest as training data. |
|     **Improved Genetic Algorithm Optimization** |
|     Define the min/max restriction on C, Epsilon, and Gamma. |
|     Feed the scaled training data to fit SVR. |
|     The genetic algorithm stochastically searches for parameters within restrictions. |
|     Calculate the arithmetic mean score of |
|     full training MAPE and recent 5-year training MAPE |
|     Find the best score till 30 generations. |
|     **Prediction** |
|     Predict the daily closing price of the index for the next 1 year. |
|     Evaluate the performance using test data. |

**Table 3. SVR Hyperparameter Restrictions**

| Hyperparameter | Min | Max |
| --- | --- | --- |
| C | 0.01 | 1 |
| Epsilon | 0.1 | 1 |
| RBF-SVR Gamma | 0 | 'scale'/10 |

## 5. Results and Discussion

The downloaded data from Yahoo Finance is preprocessed by removing fields other than dates and close prices using the date index as the input feature and close price as the target. Both are scaled using the standard min-max scaler equation shown in Eq. 1.

$$X_{scaled} = \frac{X - X_{min}}{X_{max} - X_{min}} \tag{1}$$

$X_{min}$ and $X_{max}$ are the minimum and maximum values of $X$, respectively, and are used to scale the prices between 0 and 1. The MAPE is used to evaluate the performance of models on the test data of global stock indices. Since MAPE provides the error as a percentage, it is suitable for comparing stock indices with different price ranges. Python programming on Google Collab is for this study.

### 5.1. NIFTY

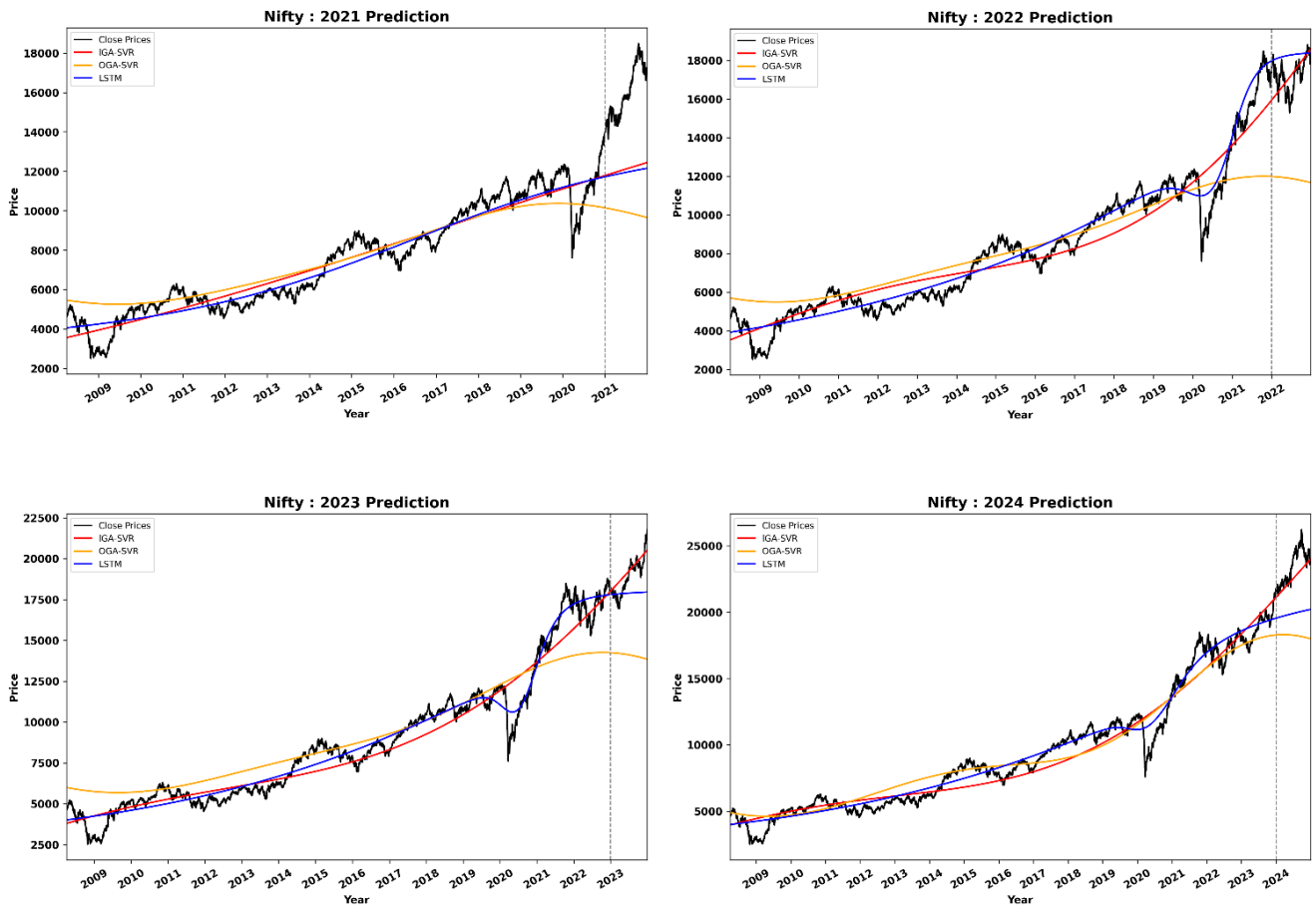

Figure 7. NIFTY Predictions

Figure 7 visualizes the Nifty prediction of three models, namely LSTM, OGA-SVR, and IGA-SVR, for four separate one-year prediction windows on testing data. Table 4 presents the forecasting performance, which also includes the average Mean Absolute Percentage Error (MAPE) across all years and the corresponding execution times. Further, Table 4 also includes hyperparameters selected by the GA each year (C, epsilon, gamma) for IGA-SVR and

Author at: Delhi School of Management, Delhi Technological University, Delhi 110042, India
E-mail addresses: mohitbeniwal@dtu.ac.in (M. Beniwal)

OGA-SVR. The LSTM achieves a good average MAPE of 13.04%, but it is the slowest among all the models with an execution time of 277 seconds. OGA-SVR is the quickest among the three models, taking only 4 seconds on average; however, it has the highest MAPE of 28.99%, making it the least accurate model. IGA-SVR achieves the lowest average MAPE of 8.7%, outperforming both LSTM and OGA-SVR. The average execution time of IGA-SVR is twice that of OGA-SVR but still a fraction of LSTM's execution time. Further year-by-year analysis shows that IGA-SVR consistently achieves lower MAPE values, especially from 2022 onward, where its MAPE dropped below 4%. These findings strongly indicate that IGA-SVR has robust predictive accuracy for Nifty over multiple years of yearly prediction windows.

**Table 4. NIFTY**

| Year | Metric | LSTM | OGA-SVR | IGA-SVR |
|---|---|---|---|---|
| | GA Selected Parameters (C,ε,g) | | 0.001, 0.1, 'scale' | 0.95, 0.1, 0.22 |
| 2021 | MAPE% | 24.97 | 37.62 | 23.91 |
| | Execution Time (in s) | 235 | 4 | 6 |
| | GA Selected Parameters (C,ε,g) | | 0.001, 0.1, 'scale' | 1, 0.1, 1.37 |
| 2022 | MAPE% | 6.25 | 31.04 | 3.99 |
| | Execution Time (in s) | 281 | 3 | 8 |
| | GA Selected Parameters (C,ε,g) | | 0.001, 0.1, 'scale' | 1, 0.1, 0.92 |
| 2023 | MAPE% | 5.91 | 25.13 | 2.97 |
| | Execution Time (in s) | 287 | 4 | 8 |
| | GA Selected Parameters (C,ε,g) | | 0.01, 0.1, 'scale' | 1, 0.11, 1.35 |
| 2024 | MAPE% | 15.02 | 22.15 | 3.93 |
| | Execution Time (in s) | 306 | 4 | 9 |
| Avg | Average MAPE% | 13.04 | 28.99 | 8.7 |
| | Average Execution Time (in s) | 277 | 4 | 8 |

### 5.2. Dow Jones Industrial Average (DJI)

Figure 8 visualizes the DJI prediction of three models. Table 5 presents the forecasting performance. Each year, IGA-SVR showed better results than LSTM. Considering overall performance, IGA-SVR demonstrated superior performance in terms of MAPE compared to LSTM and OGA-SVR. The execution time of IGA-SVR is again twice that of OGA-SVR but significantly lower than LSTM. Overall, IGA-SVR is nearly 45% more accurate than OGA-SVR. Although OGA-SVR is the least accurate, for 2022, it gave the best result compared to OGA-SVR and LSTM.

**Table 5. DJI**

| Year | Metric | LSTM | OGA-SVR | IGA-SVR |
|---|---|---|---|---|
| | GA Selected Parameters (C,ε,g) | | 0.1, 0.1, 'scale' | 0.94, 0.1, 0.35 |
| 2021 | MAPE% | 15.63 | 17.76 | 14.22 |
| | Execution Time (in s) | 241 | 3 | 6 |
| | GA Selected Parameters (C,ε,g) | | 0.01, 0.1, 'scale' | 1, 0.1, 1.37 |
| 2022 | MAPE% | 10.48 | 4.62 | 8.7 |
| | Execution Time (in s) | 266 | 3 | 5 |
| | GA Selected Parameters (C,ε,g) | | 0.01, 0.1, 'scale' | 0.01, 0.1, 1.36 |
| 2023 | MAPE% | 3.69 | 4.68 | 2.16 |
| | Execution Time (in s) | 290 | 4 | 9 |
| | GA Selected Parameters (C,ε,g) | | 0.001, 0.1, 'scale' | 0.99, 0.1, 0.34 |
| 2024 | MAPE% | 10.44 | 25.99 | 4.17 |
| | Execution Time (in s) | 304 | 5 | 11 |
| Avg | Average MAPE% | 10.06 | 13.26 | 7.31 |
| | Average Execution Time (in s) | 275 | 4 | 8 |

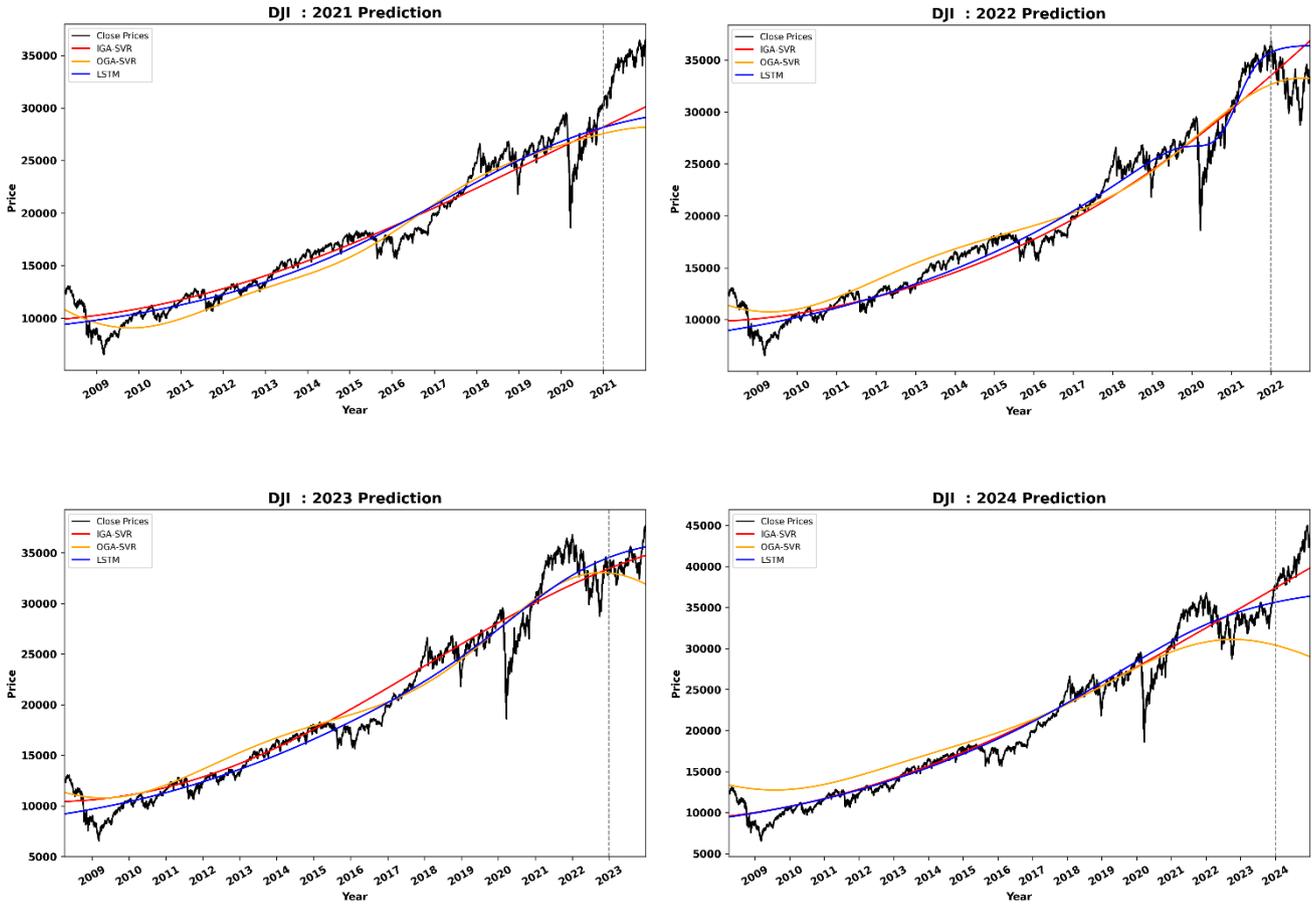

Figure 8. DJI Predictions

### 5.3. DAX Performance Index (DAX)

Unlike Nifty and DJI, the DAX shows mixed results when comparing the performance of IGA-SVR and LSTM. While IGA-SVR performed better than LSTM in 2021 and 2024, LSTM outperformed IGA-SVR in 2022 and 2023. However, when considering the overall average across all years, IGA-SVR achieved the lowest MAPE of 10.73%, indicating a competitive performance compared to LSTM with a MAPE of 12.09%. OGA-SVR continues to show weakness in forecasting the long-term prices of DAX showing the highest MAPE of 18.14% among all the three models. Table 6 presents the results and Figure 9 visualizes them.

Author at: Delhi School of Management, Delhi Technological University, Delhi 110042, India
E-mail addresses: mohitbeniwal@dtu.ac.in (M. Beniwal)

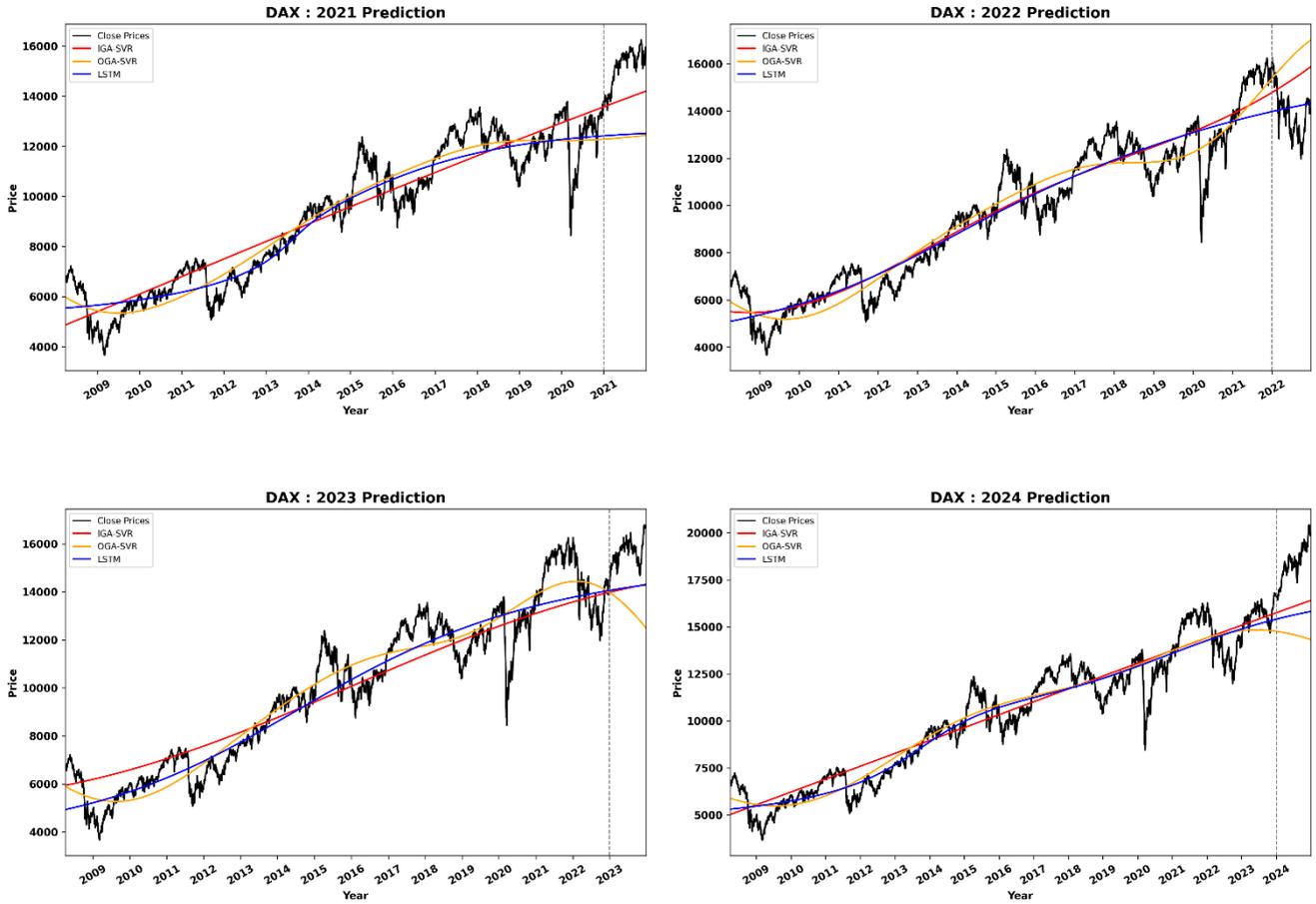

**Figure 9.** DAX Predictions

**Table 6. DAX**

| Year | Metric | LSTM | OGA-SVR | IGA-SVR |
|------|--------|------|---------|---------|
| | GA Selected Parameters (C,ε,g) | | 0.1, 0.1, 'scale' | 0.97, 0.1, 0.05 |
| **2021** | MAPE% | 17.85 | 18.61 | 8.52 |
| | Execution Time (in s) | 227 | 3 | 14 |
| | GA Selected Parameters (C,ε,g) | | 0.1, 0.1, 'scale' | 1, 0.1, 1.37 |
| **2022** | MAPE% | 6.09 | 18.25 | 12.26 |
| | Execution Time (in s) | 278 | 4 | 11 |
| | GA Selected Parameters (C,ε,g) | | 0.1, 0.1, 'scale' | 0.19, 0.20, 1.21 |
| **2023** | MAPE% | 9.45 | 15.04 | 9.62 |
| | Execution Time (in s) | 249 | 3 | 9 |
| | GA Selected Parameters (C,ε,g) | | 0.01, 0.1, 'scale' | 0.59, 0.1, 0.03 |
| **2024** | MAPE% | 14.98 | 20.67 | 12.51 |
| | Execution Time (in s) | 309 | 4 | 18 |
| **Avg** | Average MAPE% | 12.09 | 18.14 | 10.73 |
| | Average Execution Time (in s) | 266 | 4 | 13 |

*5.4. Nikkei 225 (N225)*

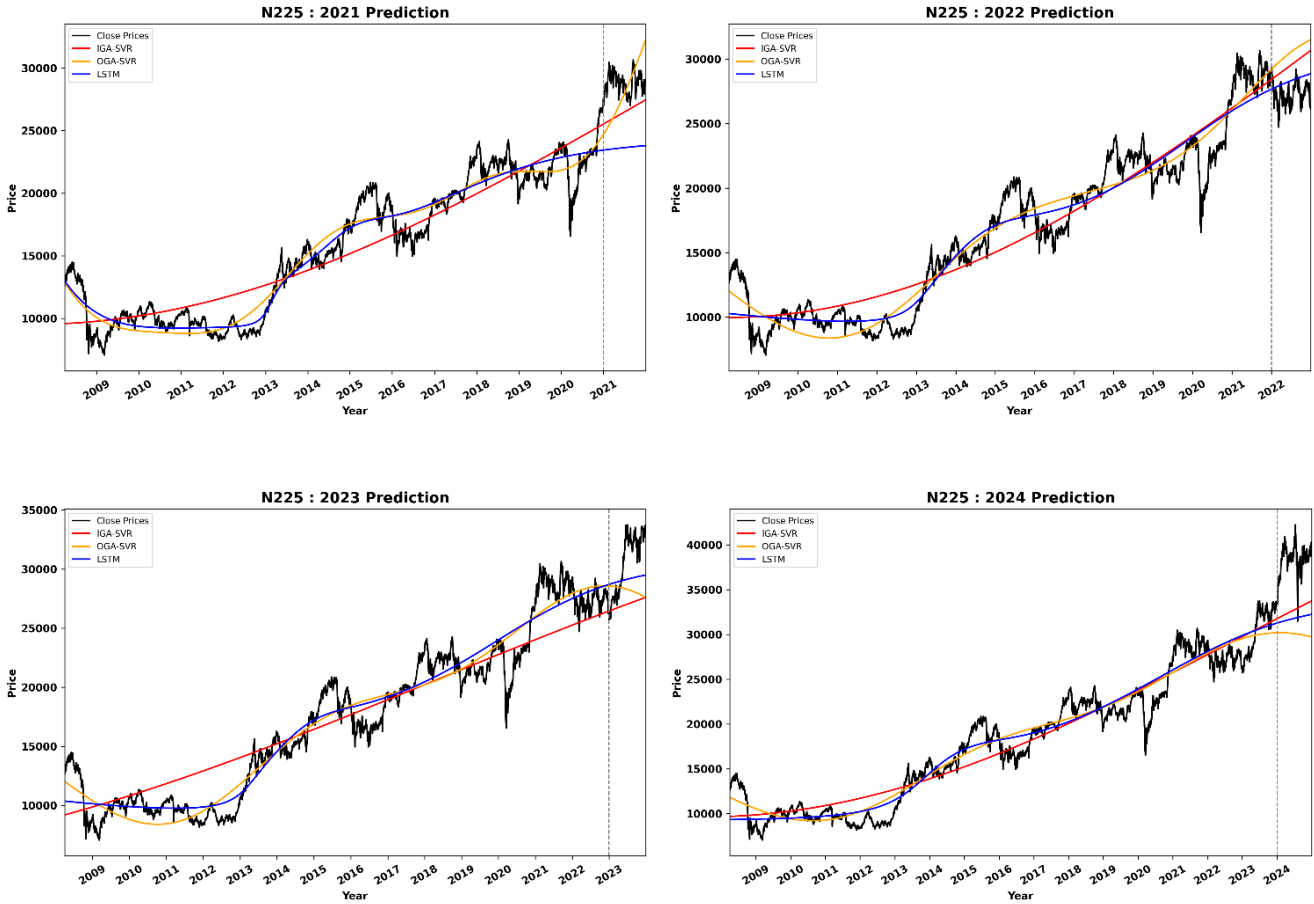

Figure 10. N225 Predictions

Figure 10 visualizes the results of N225. LSTM incurred the highest computational cost like previous results but again showed a competitive prediction performance for N225. It performed the worst in 2021 but achieved the best prediction accuracy in 2022 and 2023. Although IGA-SVR did not surpass LSTM in 2022 and 2023, it performed better than LSTM in 2021 and 2024. Overall, IGA-SVR again showed a competitive performance with a MAPE of 10.67%, outperforming both LSTM (MAPE 11.78%) and OGA-SVR (MAPE 12.71%). OGA-SVR outperformed both LSTM and IGA-SVR in 2021 only. On average in N225, OGA-SVR performance is closer to that of LSTM and IGA-SVR while being the fastest in terms of computational cost. Considering the overall performance for the N225 index, IGA-SVR demonstrated slightly superior predictive performance compared to both LSTM and OGA-SVR. Table 7 presents the result of N225.

Author at: Delhi School of Management, Delhi Technological University, Delhi 110042, India
E-mail addresses: mohitbeniwal@dtu.ac.in (M. Beniwal)

**Table 7. N225**

| Year | Metric | LSTM | OGA-SVR | IGA-SVR |
|------|--------|------|---------|---------|
| | GA Selected Parameters (C,ε,g) | | 1, 0.1, 'scale' | 0.99, 0.11, 0.21 |
| **2021** | MAPE% | 17.99 | 7.42 | 8.05 |
| | Execution Time (in s) | 245 | 4 | 18 |
| | GA Selected Parameters (C,ε,g) | | 0.1, 0.1, 'scale' | 0.91, 0.10, 0.23 |
| **2022** | MAPE% | 4.34 | 11.97 | 8.59 |
| | Execution Time (in s) | 270 | 3 | 18 |
| | GA Selected Parameters (C,ε,g) | | 0.1, 0.1, 'scale' | 0.05, 0.20, 0.29 |
| **2023** | MAPE% | 7.74 | 9.87 | 11.47 |
| | Execution Time (in s) | 268 | 4 | 16 |
| | GA Selected Parameters (C,ε,g) | | 0.01, 0.1, 'scale' | 0.69, 0.1, 0.3 |
| **2024** | MAPE% | 17.06 | 21.58 | 14.56 |
| | Execution Time (in s) | 299 | 4 | 17 |
| **Avg** | Average MAPE% | 11.78 | 12.71 | 10.67 |
| | Average Execution Time (in s) | 271 | 4 | 17 |

## 5.5. Shanghai Stock Exchange Composite Index (SSE)

Table 8 presents the year-wise performance of LSTM, OGA-SVR, and IGA-SVR on the SSE index. The overall performance of IGA-SVR on SSE was again better than that of LSTM. IGA-SVR outperformed LSTM in all years except 2022. The OGA-SVR model selected the value of parameter C, ranging from 3 to 45, resulting in the model being highly sensitive and performing nearly 120% worse than IGA-SVR for long-term forecasting of SSE. This highlights the importance of restricting the upper bound of C and other parameters of IGA-SVR, which helped improve stability and accuracy. Figure 11 visualizes the results of SSE.

**Table 8. SSE**

| Year | Metric | LSTM | OGA-SVR | IGA-SVR |
|------|--------|------|---------|---------|
| | GA Selected Parameters (C,ε,g) | | 10, 0.11, 'scale' | 0.46, 0.19, 0.79 |
| **2021** | MAPE% | 12.38 | 9.5 | 3.05 |
| | Execution Time (in s) | 223 | 3 | 17 |
| | GA Selected Parameters (C,ε,g) | | 3.04, 0.1, 'scale' | 0.35, 0.21, 0.39 |
| **2022** | MAPE% | 4.36 | 28.62 | 14.5 |
| | Execution Time (in s) | 275 | 4 | 12 |
| | GA Selected Parameters (C,ε,g) | | 38.35, 0.13, 'scale' | 0.76, 0.15, 1.32 |
| **2023** | MAPE% | 9.46 | 6.19 | 5.66 |
| | Execution Time (in s) | 290 | 5 | 20 |
| | GA Selected Parameters (C,ε,g) | | 44.80, 0.15, 'scale' | 0.19, 0.29, 0 |
| **2024** | MAPE% | 8.41 | 19.81 | 5.39 |
| | Execution Time (in s) | 294 | 5 | 39 |
| **Avg** | Average MAPE% | 8.65 | 16.03 | 7.15 |
| | Average Execution Time (in s) | 271 | 4 | 22 |

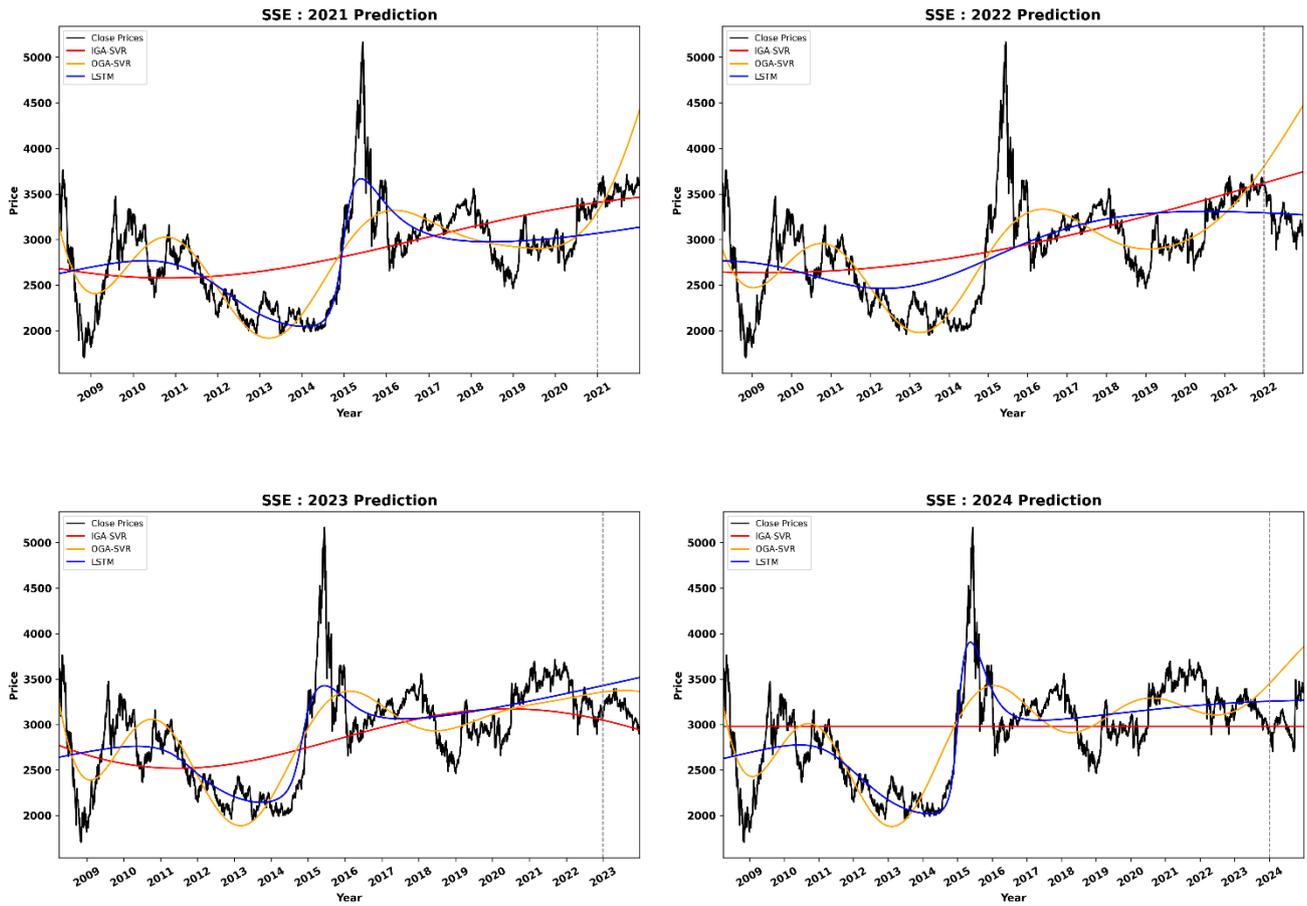

Figure 11. SSE Predictions

### 5.6. Consolidated result

Table 9 presents the consolidated results of all three models. In 2021, IGA-SVR performed significantly better, with a MAPE approximately 35% and 36% lower than that of LSTM and OGA-SVR, respectively. In 2022, LSTM performance improved significantly, and it outperformed both IGA-SVR and OGA-SVR. The MAPE of LSTM and IGA-SVR models showed improvements in 2022 compared to 2021, but OGA-SVR performance remained like 2021. In 2023, the performance of both IGA-SVR and OGA-SVR was improved compared to the previous year, with IGA-SVR outperforming the other models. However, the difference in MAPE between LSTM and IGA-SVR was less than 1%, and both models performed considerably better than OGA-SVR in 2023. In 2024, all models showed a big decline in performance, while MAPE of IGA-SVR increased by less than 2%. This highlights IGA-SVR's robustness in prediction accuracy across all years. Overall, IGA-SVR is superior for long-term forecasting with a reduction in MAPE by 19.87% compared to LSTM and 50.03% compared to OGA-SVR. LSTM also performed moderately, while OGA-SVR was consistently the weakest performer when evaluated over the four-year period from 2021 to 2024 on test data. The execution time for LSTM was approximately 20 times higher than that of IGA-SVR. This is a significantly higher computational cost compared to the more efficient and accurate IGA-SVR. This can be a concern for some investors or institutions who are seeking a faster but still accurate model.

Author at: Delhi School of Management, Delhi Technological University, Delhi 110042, India
E-mail addresses: mohitbeniwal@dtu.ac.in (M. Beniwal)

**Table 9. Consolidated**

| Year | Metric | LSTM | OGA-SVR | IGA-SVR |
|------|--------|------|---------|---------|
| 2021 | MAPE% | 17.76 | 18.18 | 11.55 |
| | Execution Time (in s) | 234 | 3 | 12 |
| 2022 | MAPE% | 6.3 | 18.9 | 9.61 |
| | Execution Time (in s) | 274 | 3 | 11 |
| 2023 | MAPE% | 7.25 | 12.18 | 6.38 |
| | Execution Time (in s) | 277 | 4 | 12 |
| 2024 | MAPE% | 13.18 | 22.04 | 8.11 |
| | Execution Time (in s) | 302 | 4 | 19 |
| Overall | Average MAPE% | 11.12 | 17.83 | 8.91 |
| | Average Execution Time (in s) | 272 | 4 | 14 |

Table 10 presents the performance of IGA-SVR across all indices on a year-wise basis, along with the overall mean performance across indices and years. The table visualizes the performance using a color-coded representation. Green color indicates the best performance among the three models, orange represents moderate performance, which means that IGA-SVR ranked second among the three models, and red indicates that IGA-SVR has the highest MAPE among the three models. The results strongly suggest that for Nifty, IGA-SVR is the most suitable model across all years. Similarly, for DJI and SSE, IGA-SVR outperformed the other models in all years except for 2022. For DAX, IGA-SVR achieved the best performance on the test data in 2021 and 2024, while its performance was moderate in 2022 and 2023. The only exception was for N225 in 2023, where IGA-SVR performed less than both LSTM and OGA-SVR. Moreover, IGA-SVR performed moderately in 2021 and 2022. Considering overall year-wise performance across indices, IGA-SVR performed the best in all years except 2022. Further, when evaluating index-wise performance across years, IGA-SVR strongly emerged as the superior model for all indices. The results clearly suggest the superiority of the IGA-SVR model.

**Table 10. IGA-SVR Performance**

### 5.7. Discussion and Managerial Insights

LSTM is an excellent model specifically designed for sequential data. Financial time series data is also a sequence of prices, making LSTM naturally suitable for such applications. LSTM has built-in mechanisms to capture temporal dependencies, making it effective for modeling time series data. Several researchers have also found it suitable for financial time series forecasting. Investors and managers can leverage LSTM for long-term forecasting. Furthermore, stacking multiple layers of LSTM makes it a deep learning model capable of automatically extracting features without requiring explicit feature engineering. However, multi-layer LSTM, as a deep learning model, requires a large amount of training data, resulting in high computational costs. In contrast, the novel Improved Genetic Algorithm-optimized Support Vector Regression (IGA-SVR) demonstrated superior forecasting performance compared to the state-of-the-art deep learning LSTM model while requiring only approximately 5% of the execution time compared to 100 epochs of two layers LSTM model. To the best of our knowledge, considering the overall performance of IGA-SVR across all indices and years, IGA-SVR emerges as the most effective model for long-term stock index forecasting, offering better accuracy than LSTM with minimal computational cost. Therefore, managers and investors are recommended to adopt the IGA-SVR model as a viable alternative to LSTM to achieve both improved prediction accuracy and significantly lower computational cost. Further, IGA-SVR, when used for multiple years of daily price prediction up to a year, is capable of giving better accuracy for a global stock indices.

## 6. Conclusion and future recommendations

Predicting the stock market is an arduous task, as stock prices can move abruptly based on unseen future events. Still, it is worthwhile to have generalized forecasting, even with lower accuracy, to guide investors and financial institutions. Long-term forecasting of stock prices further makes the task daunting. LSTM has a great ability to handle sequences of data. In this study, LSTM and OGA-SVR, examined by Beniwal et al. [45], were taken as baseline models for comparison. This study proposed Improved Genetic Algorithm Optimized Support Vector Regression (IGA-SVR) and comprehensively evaluated the proposed model on five major indices of the top five GDPs of the world, namely the Dow Jones Industrial Average (DJI) from the USA, the Shanghai Stock Exchange Composite Index (SSE) from China, the DAX performance Index (DAX) from Germany, the Nikkei 225 (N225) from Japan, and Nifty from India. The study conducted an in-depth analysis to evaluate the performance of models using testing data for each year's prediction from 2021 to 2024.

The proposed IGA-SVR consistently outperformed both LSTM and OGA-SVR across the majority of years for all five indices, achieving an overall average MAPE of 8.91%. In comparison, LSTM and OGA-SVR achieved an overall average MAPE of 11.12% and 17.83%, respectively. This represents a 19.87% improvement over LSTM and a 50.03% improvement over OGA-SVR, which is a considerable difference in forecasting a noisy stock market. Another important achievement of the proposed IGA-SVR model is that its execution time was approximately 5% of the execution time of LSTM. This highlights the superior computational efficiency in addition to improved accuracy. These attributes make a strong case for IGA-SVR as a viable and effective alternative model for the long-term forecasting of global stock indices. This study also employs an intuitive training methodology that uses the arithmetic mean of the score of the entire historical training dataset while simultaneously extracting a recent score based on the last five years of training data. The genetic algorithm uses this score to tune those hyperparameters of the SVR model that consider historical trends and recent trends in the stock index. Computation cost was reduced for GA by restricting the solutions of hyperparameters to prevent the model from becoming overly sensitive to short-term noise. Thus, the improved model has a better selection of hyperparameters and is more adaptive to changing long-term trends with lower computational costs. In contrast, OGA-SVR employed a rolling forward validation approach that only utilized recent data, causing the model to forget long-term trends. Additionally, OGA-SVR lacked effective constraints on hyperparameter selection and ignored the optimization of gamma, which controls model sensitivity and influence of individual data points.

In summary, IGA-SVR demonstrated high effectiveness and computational efficiency for long-term forecasting of global stock indices, establishing itself as a viable alternative to state-of-the-art deep learning models such as LSTM for investors and financial institutions. Although the model was evaluated using daily frequency data, it can be extended to lower-frequency data for multi-step predictions, which could also be valuable for traders. Additionally, the model can potentially be adapted to enhance investment returns, as the price levels predicted by the model could help investors identify value opportunities below the forecasted price. Furthermore, IGA-SVR can play a role in developing trading strategies and risk management systems by providing reliable long-term forecasts. Future research can also explore the applications of the proposed model to sectoral indices or individual stocks.

One of the limitations of this study is that the restrictions imposed on the hyperparameters of SVR were manually identified by trial and error. Future research could explore data-driven approaches to constrain the hyperparameter search space based on the training data itself. Further, as global indices tend to exhibit lower volatility compared to individual stocks, it remains an open question whether IGA-SVR will maintain its superior performance when applied to more volatile price movements in individual stocks.

### Declaration of Competing Interest

The authors state that there are no financial or personal conflicts of interest that could have influenced the findings reported in this paper.

### Data Availability

All data used in this study are publicly available on Yahoo Finance.

Author at: Delhi School of Management, Delhi Technological University, Delhi 110042, India
E-mail addresses: mohitbeniwal@dtu.ac.in (M. Beniwal)

**Appendix A**

| Abbreviation | Explanation |
|---|---|
| LSTM | Long Short-Term Memory |
| RNN | Recurrent Neural Network |
| SVR | Support Vector Regression |
| ML | Machine Learning |
| ARIMA | Auto-Regressive Integrated Moving Average |
| SVM | Support Vector Machine |
| GS | Grid-Search |
| GA | Genetic Algorithms |
| AI | Artificial Intelligence |
| PSO | Particle Swarm Optimization |
| OGA-SVR | Optimized Genetic Algorithm Support Vector Regression |
| IGA-SVR | Improved Genetic Algorithm Optimized Support Vector Regression |
| MAPE | Mean Absolute Percentage Error |
| DJI | Dow Jones Industrial Average |
| DAX | DAX Performance Index |
| N225 | Nikkei 225 |
| SSE | Shanghai Stock Exchange Composite Index |

Author at: Delhi School of Management, Delhi Technological University, Delhi 110042, India
E-mail addresses: mohitbeniwal@dtu.ac.in (M. Beniwal)

Author at: Delhi School of Management, Delhi Technological University, Delhi 110042, India
E-mail addresses: mohitbeniwal@dtu.ac.in (M. Beniwal)